\documentclass[amsmath,amssymb,prl,reprint,superscriptaddress]{revtex4-1}
\usepackage{graphicx}% Include figure files
\usepackage{dcolumn}% Align table columns on decimal point
\usepackage{bm}% bold math
\usepackage{natbib}

\usepackage{color} 
\newcommand{\dv}{\mathrm{d}}
\newcommand{\dVdI}{$\frac{\dv V}{\dv I}$}

\setcitestyle{super}

\begin{document}

\title{Andreev reflection in s-type superconductor proximized 3D topological insulator}
\author{E.S.~Tikhonov}
\affiliation{Institute of Solid State Physics, Russian Academy of
Sciences, 142432 Chernogolovka, Russian Federation}
\affiliation{Moscow Institute of Physics and Technology, Dolgoprudny, 141700 Russian Federation}
\author{D.V.~Shovkun} 
\affiliation{Institute of Solid State Physics, Russian Academy of
Sciences, 142432 Chernogolovka, Russian Federation}
\affiliation{Moscow Institute of Physics and Technology, Dolgoprudny, 141700 Russian Federation}
\author{M. Snelder}
\affiliation{MESA+ Institute for Nanotechnology, University of Twente, Enschede, the Netherlands}
\author{M.P. Stehno}
\affiliation{MESA+ Institute for Nanotechnology, University of Twente, Enschede, the Netherlands}
\author{Y. Huang}
\affiliation{Van der Waals - Zeeman institute, University of Amsterdam, the Netherlands.}
\author{M.S. Golden}
\affiliation{Van der Waals - Zeeman institute, University of Amsterdam, the Netherlands.}
\author{A.A. Golubov}
\affiliation{MESA+ Institute for Nanotechnology, University of Twente, Enschede, the Netherlands}
\affiliation{Moscow Institute of Physics and Technology, Dolgoprudny, 141700 Russian Federation}
\author{ A. Brinkman}
\affiliation{MESA+ Institute for Nanotechnology, University of Twente, Enschede, the Netherlands}
\author{V.S.~Khrapai} 
\affiliation{Institute of Solid State Physics, Russian Academy of
Sciences, 142432 Chernogolovka, Russian Federation}
\affiliation{Moscow Institute of Physics and Technology, Dolgoprudny, 141700 Russian Federation}

\begin{abstract}
 We investigate transport and shot noise in lateral N-TI-S contacts, where N is a normal metal, TI is a Bi-based three dimensional  topological insulator (3D TI), and S is an s-type superconductor. In the normal state, the devices are in the elastic diffusive transport regime, as demonstrated by a nearly universal value of the shot noise Fano factor $F_{\rm N}\approx1/3$ in magnetic field and in a reference normal metal contact. In the absence of magnetic field, we identify the Andreev reflection (AR) regime, which gives rise to the effective charge doubling in shot noise measurements. Surprisingly, the Fano factor $F_{\rm AR}\approx0.22\pm0.02$ is considerably reduced in the AR regime compared to $F_{\rm N}$, in contrast to previous AR experiments in normal metals and semiconductors. We suggest that this effect is related to a finite thermal conduction of the proximized, superconducting TI owing to a residual density of states at low energies.
\end{abstract}

\maketitle

A surface state of a three-dimensional topological insulator (3D TI) is a unique example of a spin-orbit coupled and symmetry protected  conductor~\cite{Hasan2010}. Similar to graphene, in 3D TI the surface electronic states are massless Dirac fermions. Unlike in graphene, however, a single Dirac cone is lacking spin and valley degeneracies. That makes a 3D TI an intriguing candidate for the realization of a solid state two-dimensional topological superconductor~\cite{Qi2011}. As originally proposed by Fu and Kane~\cite{Fu2008}, p-wave like superconducting correlations are expected to emerge via proximity coupling the 3D TI to a conventional s-type superconductor (S). This gives rise to symmetry protected Majorana zero modes bound at vortices or at the boundaries of various hybrid structures~\cite{Tanaka2009,Alicea2012}. Emerging zero modes are predicted to have a strong impact on the low energy physics, modifying Andreev reflection (AR) at the interface with a normal metal~\cite{Beenakker2011}, affecting the edge conductance distribution~\cite{Beenakker2015}, noise~\cite{Bolech2007} and thermal transport~\cite{Akhmerov2011,Diez2014}.

Proximity induced superconductivity has been demonstrated in 3D TIs based on Bi~\cite{Zhang2011,Sacepe2011,Veldhorst2012,Wang2012_proximity,Williams2012,Cho2013,Galletti2014} and HgTe~\cite{Oostinga2013,Wiedenmann2016}  with the reported values of the induced gap on the order of a few 100~$\mu$eV. Similar gap values were recently observed via Andreev spectroscopy~\cite{Snelder2015} in quaternary BiSbTeSe compound, established as a 3D TI with negligible contribution of the bulk conduction~\cite{Ren2011,Pan2014}. In spite of these advances, the microscopic nature of the proximity induced gap in Bi-based 3D TIs remains largely unexplored. This particularly concerns the statistics of the transmission eigenvalue distribution in short junctions and the role of possible in-gap states. On this route, valuable information, often hidden in transport, can be obtained via measurements of the non-equilibrium current fluctuations --- the shot noise~\cite{Blanter2000}. Relevant for the AR, prominent examples include unusual shot noise behavior at the interface between a normal metal and superconductors with other than s-type order parameter symmetry~\cite{Zhu1999,Tanaka2000} and, more recently, shot noise detection of the thermal and charge transport via Majorana zero modes~\cite{Akhmerov2011,Diez2014,Zazunov2016}.

Here, we investigate the AR in lateral N-TI-S contacts defined on thin flakes of 3D TI $\rm Bi_{1.5}Sb_{0.5}Te_{1.7}Se_{1.3}$. By measuring the shot noise, we identify the effective charge ($q$) and the Fano factor ($F$), which characterize the random process of charge transport~\cite{Blanter2000}. We demonstrate diffusive normal transport with $q=e$ and $F_{\rm N} \simeq 1/3$ in N-TI-S contacts that are subjected to magnetic fields ($B$) that are large enough to suppresses superconductivity and in the reference N-TI-N contact. In zero magnetic field, the AR regime is characterized by the doubling of the effective charge $q=2e$ and $F_{\rm AR} = 0.22\pm0.02$. Our main observation that $F_{\rm AR}<F_{\rm N}$ is unusual compared with experiments in normal metal--~\cite{Kozhevnikov2000} and semiconductor--based\cite{Choi2005}  N-S contacts. 

%\textcolor{dickred}{We suggest that this effect is related to a finite thermal conduction of the proximized, superconducting TI owing to a residual density of states at low energies.} 

\begin{figure}[t]
\begin{center}
\vspace{0mm}
  \includegraphics[width=1\columnwidth]{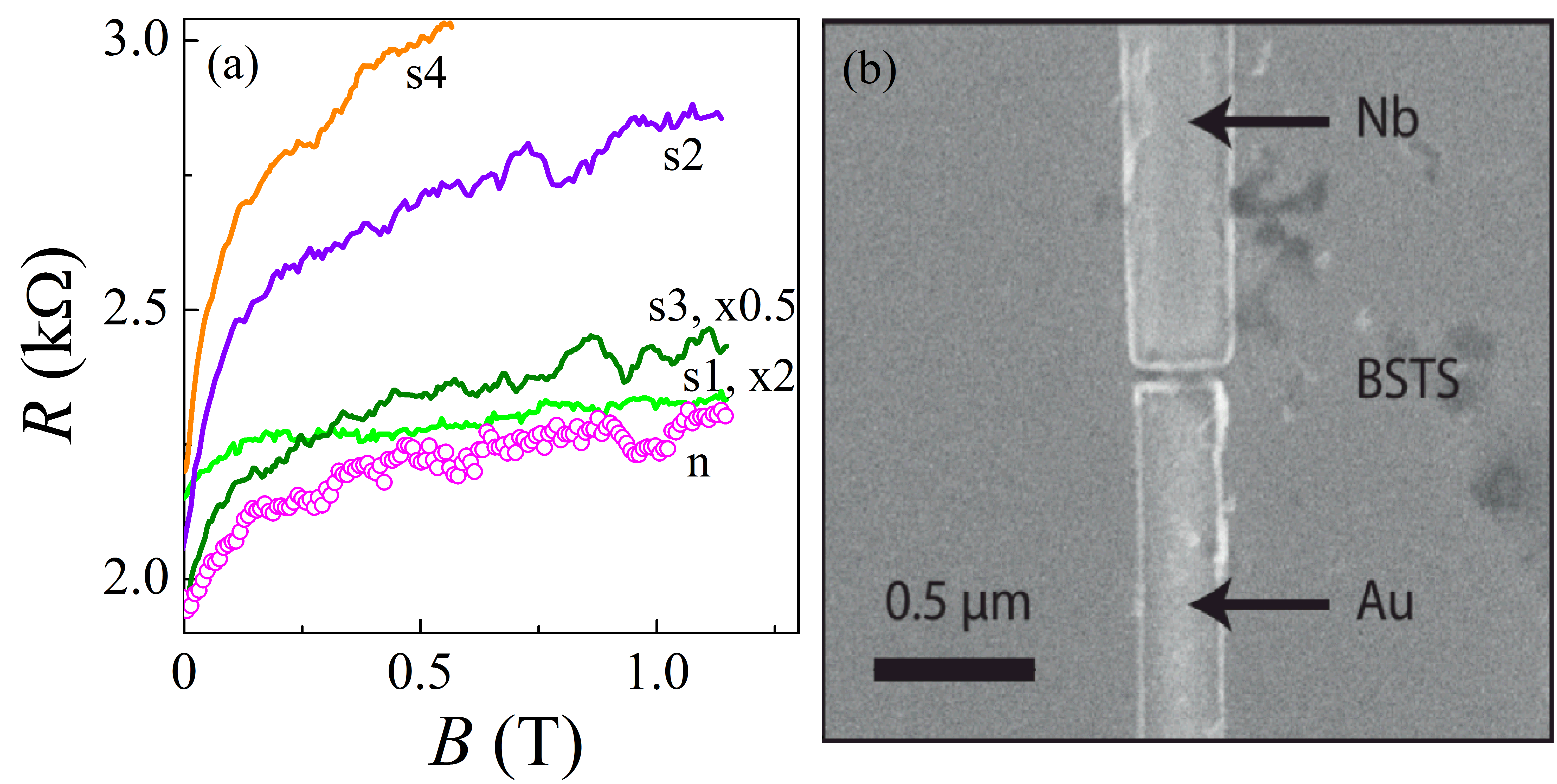}
   \end{center}
  \caption{Magnetotransport data and sample layout. On the left: linear response resistances as functions of perpendicular magnetic field for the N-TI-N device (n, symbols) and N-TI-S devices (s1-s4, lines), measured at $T=0.6\,\mathrm{K}$. For the devices s1 and s3 the data were multiplied by factors of 2 and 0.5, respectively. On the right: a scanning electron micrograph of the inner part of one of our N-TI-S devices. }\label{fig1}
\end{figure}

Our devices are based on thin mechanically exfoliated flakes of a 3D TI $\rm Bi_{1.5}Sb_{0.5}Te_{1.7}Se_{1.3}$ crystal placed on a $\rm Si/SiO_2$ substrate. The size of the flakes is $\sim 2-4\,\mu$m with a thickness in the range of $80-200\,$nm. The typical carrier densities and mean-free paths in our devices are on the order of $\rm 10^{13}\,cm^{-2}$ and  $l\sim 10$\,nm, respectively. A scanning electron micrograph of a typical N-TI-S device is shown in Fig.~\ref{fig1} (on the r.h.s.). In a two step e-beam lithography process, we pattern the N electrode (60\,nm of sputter-deposited Au with a 3\,nm Ti sticking layer) followed by the S electrode (a 80\,nm  thick {Nb} film  with a critical temperature of 8.4\,K). A 15\,s long 300\,V RF Ar plasma etch was carried out in-situ prior to the sputter-deposition in order to improve the quality of the interfaces. The width of the electrodes is about 300\,nm, and the electrode spacing, $L\approx50$\,nm, is limited by the lithography resolution. These dimensions allowed us to minimize inelastic scattering without compromising a well-defined geometry of the conduction channels. Our experiment is performed in the diffusive limit, $L\gg l$, such that residual inhomogeneities of the current distribution, caused by unintentional roughness of the edges of the metallic electrodes and their mutual random misalignment are not important.
Altogether we studied four N-TI-S devices and one N-TI-N device. The measurements were performed in a $^3$He refrigerator at bath temperatures ($T$) in the range of 0.6\,K-5\,K. Two-terminal differential resistance ($R$) data were obtained with a lock-in measurement and a series resistance contribution of wiring was subtracted. The shot noise was measured within a $\sim5$\,MHz frequency band around a $18$\,MHz  center frequency of a resonant tank circuit. The tank circuit consisted of a $3.2\,\mu$H hand-wound inductor, the 25\,pF capacitance of a coaxial cable and a load resistance of 10\,k$\Omega$. The setup was calibrated by means of Johnson-Nyquist thermometry. Where used, the magnetic field was directed perpendicular to the TI plane.

\begin{figure}[t]
\begin{center}
\vspace{0mm}
  \includegraphics[width=1\columnwidth]{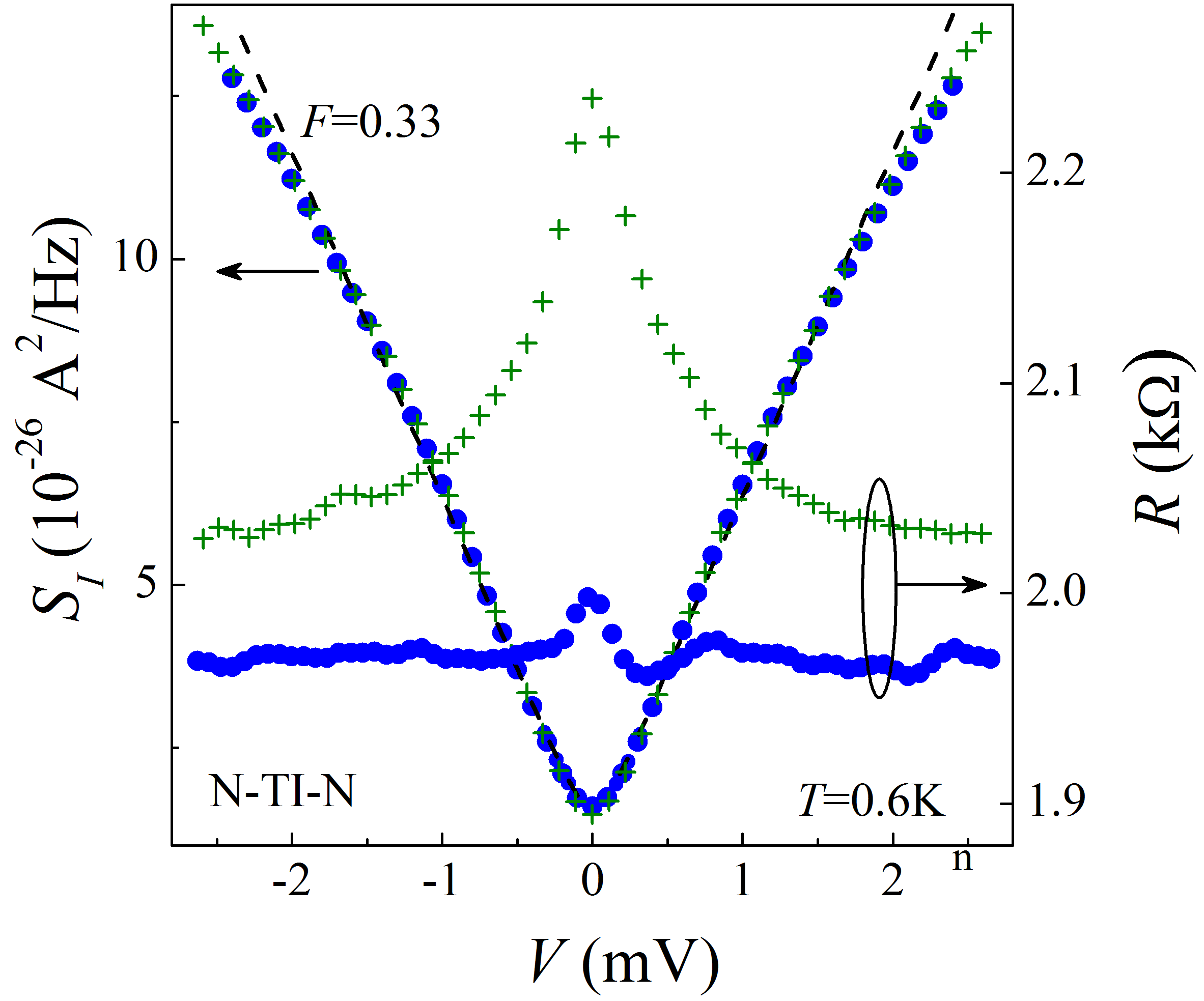}
   \end{center}
  \caption{Shot noise and transport in the reference N-TI-N device. Left axis: bias voltage dependence of the measured shot noise spectral density at $B=0\,\mathrm{T}$ (circles) and $B=1.2\,\mathrm{T}$ (crosses). The fit using eq.~(\ref{noise_expression}) with $F=0.33$ is shown by the dashed line. Right axis: differential resistance $R=$\dVdI as a function of the bias voltage across the device at $B=0$  (circles) and $B=1.2\,\mathrm{T}$ (crosses).}\label{fig2}
\end{figure}

\begin{figure}[t]
\begin{center}
\vspace{0mm}
  \includegraphics[width=0.8\columnwidth]{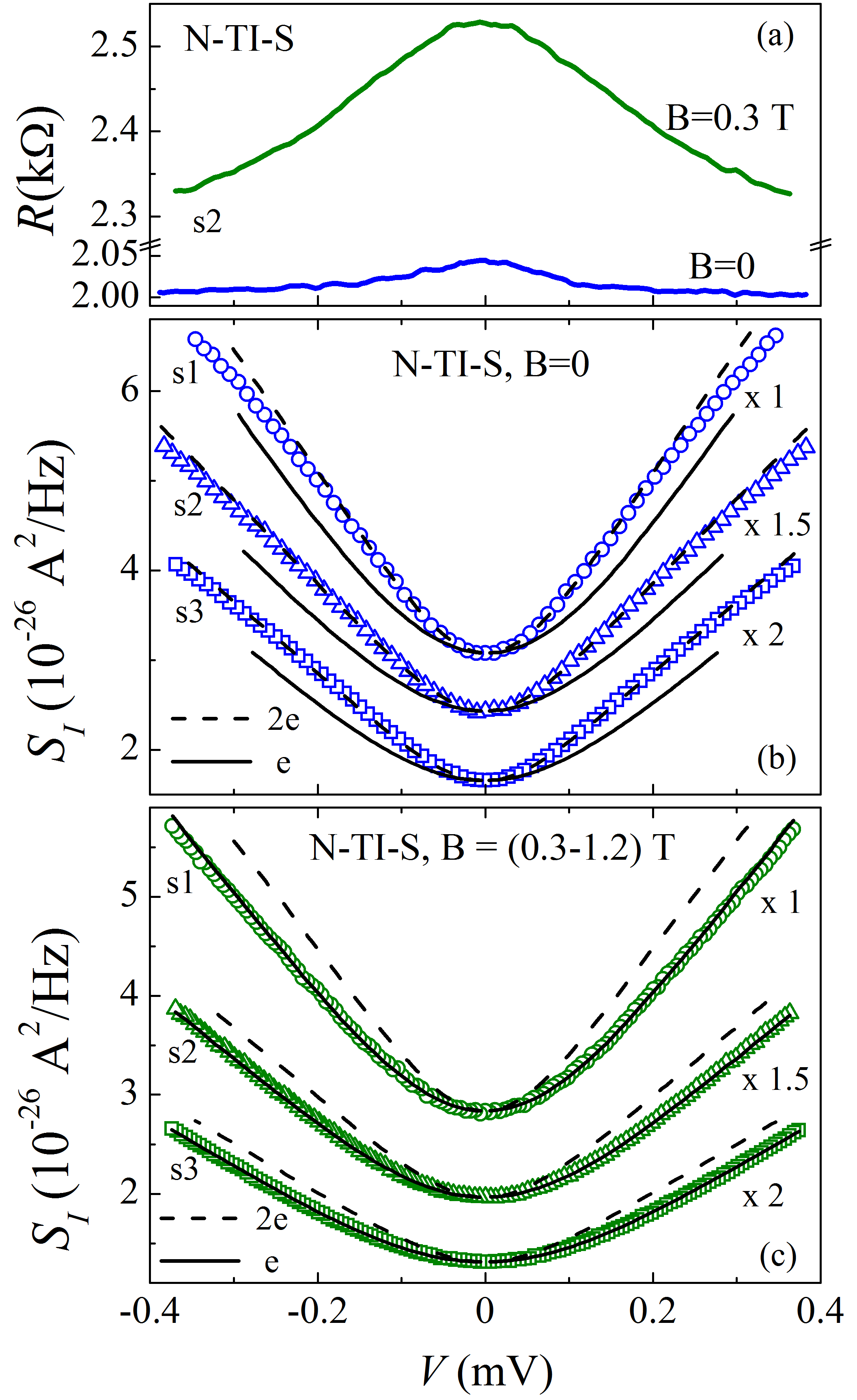}
   \end{center}
  \caption{Transport and shot noise in N-TI-S devices. (a)~Typical bias voltage dependence of the differential resistance measured in a N-TI-S device s2 at $B=0\,\mathrm{T}$ and $B=0.3\,\mathrm{T}$. (b)~Measured shot noise spectral density (symbols) as a function of the bias voltage across the device in zero magnetic field. The fits using eq.~(\ref{noise_expression}) with $q=e$, $0.38\leq F \leq0.48$ and $q=2e$, $0.19\leq F\leq0.24$ are shown by solid and dashed lines, respectively. The datasets are multiplied by 1, 1.5 and 2, respectively, from top to bottom. (c)~$S_I(V)$ (symbols) in perpendicular magnetic field. The fits with $q=e$, $0.31\leq F \leq0.36$ and $q=2e$, $0.15\leq F\leq0.18$ are shown by solid and dashed lines, respectively. Different traces correspond to devices s1, s2 and s3 as marked in the figure.
}\label{fig3}
\end{figure}

In Fig.~\ref{fig1}, we plot the linear response resistances $R$ as a function of the applied magnetic field $B$ for the N-TI-S devices (s1-s4) and the reference N-TI-N device (n) at $T\approx0.6\,$K. Both the zero field resistance $R(B=0)$ and the relative magnetoresistance vary appreciably among the devices. Nevertheless, in all cases, a sizable positive magnetoresistance is found, reminiscent of a $B$-driven suppression of the weak anti-localization quantum correction in a strongly spin-orbit coupled system~\cite{Hikami1980,Altshuler1980}. The absence of a distinct difference between the N-TI-S and the N-TI-N data is a strong indication that the device resistance is dominated by the disordered 3D TI surface. This conjecture is confirmed by the shot noise experiments below.

Diffusive transport behavior ($L\gg l$) is showcased in Fig.~\ref{fig2} (left axis), where we plot the shot noise spectral density $S_I$ (symbols) as a function of the bias voltage $V$ for the N-TI-N device at $T\approx 0.6\,$K. The $B=0$ and $B\approx1.2$\,T
data are shown with dots and crosses, respectively. In both cases, we observe conventional shot noise behavior of a metallic diffusive conductor for $|V|\leq 1.5\,$mV. Here, $S_I$ crosses over from the equilibrium Johnson-Nyquist value of $4k_BT/R$ at $V=0$ to the linear dependence, $S_I = 2qFI$, at $|V|\gg k_BT/q$, where $I$ is the bias current and $F$ is the Fano factor. This crossover is described by the standard noise expression~\cite{Blanter2000}:
\begin{equation}
	S_I=\frac{4k_BT}{R}(1-F+F\xi\coth{\xi}),\,\,\,\xi\equiv\frac{|qV|}{2k_BT},
	\label{noise_expression}
\end{equation}
 where $q$ is the effective charge of the carriers, and a linear dependence, $I=V/R$, is assumed. The corresponding best fit with $q=e$ and $F\approx 0.33$ is shown [$B=0$; dashed line]. This data is consistent with the universal [i.e., geometry independent] value $F=1/3$ found for metallic conductors in the elastic diffusive regime of transport~\cite{Nagaev1992,Beenakker_Buettiker_1992} but has not been reported in the TI surface states previously. Above $|V|\approx1.5\,$mV, the experimental data deviate from the theoretical curve, as inelastic electron-phonon scattering processes come into play and start to suppress the noise~\cite{Nagaev1992}. 

We conclude our discussion of transport data in the N-TI-N device by analyzing the differential resistance (\dVdI) plotted in Fig.~\ref{fig2}~(right axis). The variations in the differential resistance are small, which justifies our use of Eq.~(\ref{noise_expression})  and ensures accuracy within a few percent of the extracted value for $F$. Notably, while \dVdI\; exhibits a sizeable zero bias peak in magnetic field [about 10\,\% in magnitude; upper curve], a much smaller peak is observed for $B=0$ [lower curve]. This indicates the presence of an additional quantum correction to the conductance due to electron-electron interactions [EEI; Altshuler-Aronov effect], which is also supported by the observation of a weakly insulating temperature-dependence in zero magnetic field. At $B=0$, the  corrections owing to the EEI and the weak anti-localization are comparable in magnitude and roughly cancel each other in present devices~\cite{Lu2014}. 

\begin{figure}[t]
\begin{center}
\vspace{0mm}
  \includegraphics[width=0.8\columnwidth]{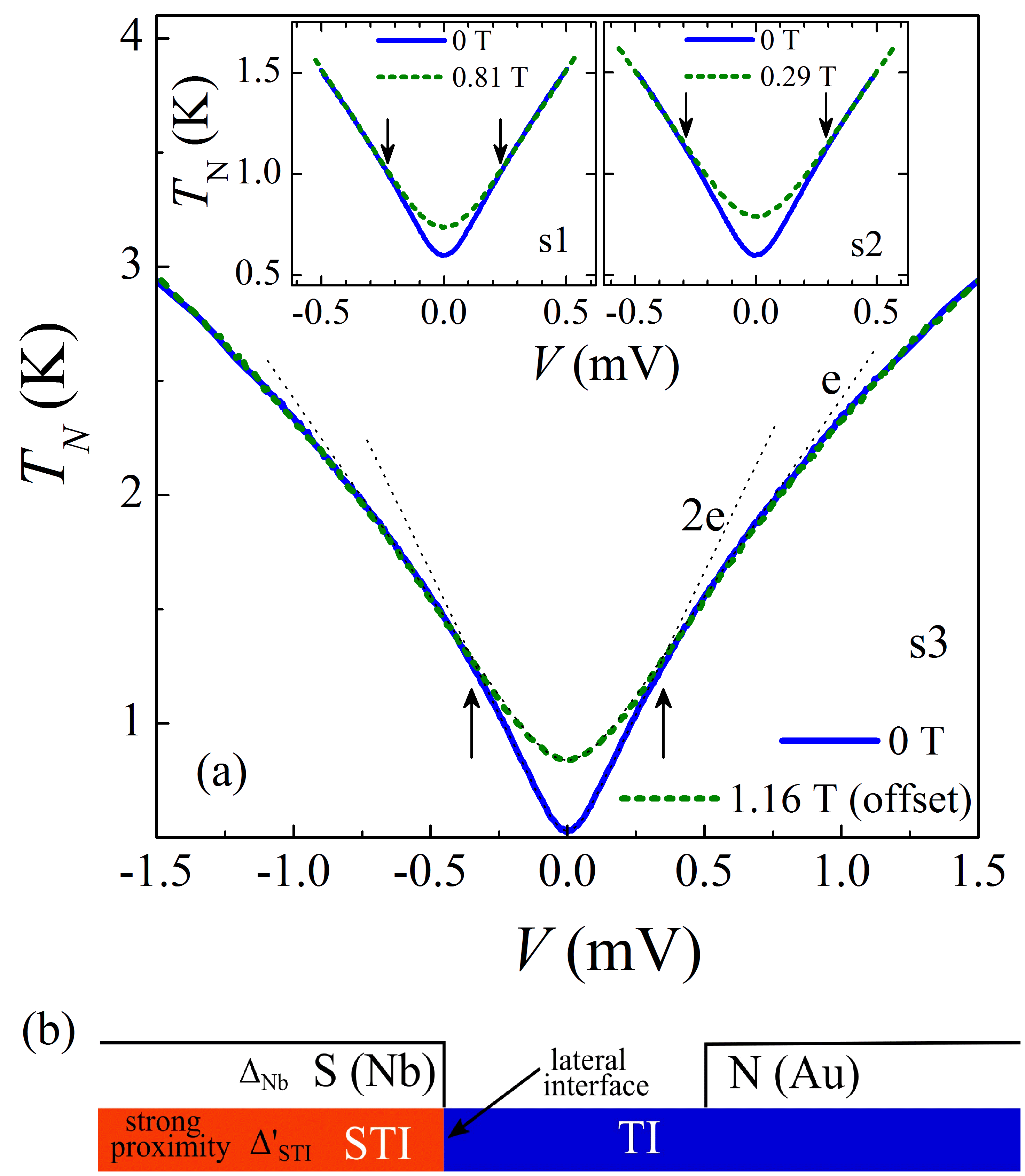}
   \end{center}  
	\caption{Finite bias $2e\rightarrow e$ transition and proximity gap. (a)~body: measured noise temperature vs $V$ in the AR regime (thick solid line) and in the normal transport regime (thick dashed line, offset by 0.32\,K) in device s3, $T\approx0.53\,$K. Corresponding $2e$ and $e$ fits are shown by thin dotted lines. The $2e\rightarrow e$ transition at the edge of the proximity gap, $|V|\approx0.35$\,mV, is marked by arrows. Note that thick solid and dashed lines closely coincide up to $|V|\approx5$\,mV, which is not shown in figure. Insets: similar $2e\rightarrow e$ transitions in the devices s1 and s2. (b)~proximity effect in a lateral N-TI-S junction. A superconducting gap $\Delta'_{\rm STI}$ opens in a strongly proximized TI (STI) beneath the Nb film. Lateral interface between the STI and normal TI, relevant for noise and transport in the AR regime, is indicated by an arrow. }\label{fig4}
\end{figure}

In the following, we discuss Andreev reflection in N-TI-S devices. In Fig.~\ref{fig3}(a), we plot typical differential resistance as a function of bias voltage, $V$, in zero field and for finite $B$. The main features of Fig.~\ref{fig3}(a), namely the smaller and narrower zero bias peaks in $B=0$ compared to the case of finite $B$-field, are similar to those in the N-TI-N device, cf. Fig.~\ref{fig2}. Such similarity persists up to 10\,mV, which is the highest bias voltage we applied, see Supplementary Material for more details~\cite{supplemental}. This further suggests that the zero bias resistance peaks are intrinsic to the 3D TI surface state. The lack of AR related resistance features allows us to estimate a transparency of the lateral interface between the normal and strongly proximized TI (Fig.~\ref{fig4}b), $\Gamma>\sqrt{l/L}\sim 0.4$, which is sufficiently high for negligible reflectionless tunneling in diffusive N-S junctions~\cite{Beenakker1997, Volkov1993,Hekking1993}. 

In Fig.~\ref{fig3}(b), we plot the bias dependence of $S_I$ for samples s1-s3 at $T\approx0.6$\,K and $B=0$. The crossover between the Johnson-Nyquist and the shot noise regime is also observed here. Yet, a closer look reveals two important distinctions compared to the N-TI-N device. Firstly, the shot noise is stronger in N-TI-S devices and, secondly, the crossover region is narrower than Eq.~(\ref{noise_expression}) predicts. This is demonstrated by the solid line fits in Fig.~\ref{fig3}(b), where we used $q=e$ and different Fano factors $0.38\leq F\leq0.48$ fixed by the  slope of each dataset in the shot noise regime. We attribute the narrowing of the crossover region to the impact of the AR at the TI-S interface, which gives rise to doubling of the effective charge in diffusive N-S contacts~\cite{Jehl2000,Kozhevnikov2000}. Using the analog of  Eq.~(\ref{noise_expression}) with effective charge~\cite{Khlus1987,Martin1996} $q=2e$ and, correspondingly, twice smaller Fano factors ($0.19\leq F\leq 0.24$), we obtained nearly perfect fits [see the dashed lines in Fig.~\ref{fig3}(b)]. As shown in Fig.~\ref{fig3}(c), the normal diffusive transport regime is restored for sufficiently high magnetic fields $B\gtrsim 1$\,T, as expected. Here, the data in all three devices (symbols) are again consistent with $q=e$ and $0.31\leq F\leq 0.36$ (solid lines), very much like in the N-TI-N contact, cf. Fig.~\ref{fig2}. By contrast, the $q=2e$ fits fail to account for the bias dependence of $S_I$ in magnetic field, see the dashed lines in Fig.~\ref{fig3}(c). 

In Fig.~\ref{fig4}(a), we plot the noise temperature, $T_N\equiv S_IR/4k_B$, as a function of $V$ in device s3 in the mV range. The $B=0$ trace [thick solid line] deviates from the $q=2e$ fit [thin dotted line] at $|V|\approx 0.35\,$mV as marked by the arrows. Remarkably, above this point, the behavior of $T_N(V)$ coincides with that of the normal transport regime. To demonstrate this, we plot the trace for applied magnetic field [$B=1.16$\,T, thick dashed line], and the $q=e$ fit [thin dotted line], both offset vertically. Here, as in previous studies~\cite{Choi2005,Das2012}, the data manifests a finite bias $2e\rightarrow e$ transition from the AR-dominated subgap transport to the normal transport regime above the gap~\cite{remark_eph}. The corresponding induced proximity gap (on the surface of the 3D TI underneath the Nb film) equals approximately $\Delta'_{\rm STI}\approx 0.35\,$meV in device s3. In the same manner, we obtained proximity gaps of $\Delta'_{\rm STI}\sim0.25-0.3\,$meV in devices s1 and s2 [see insets in Fig.~\ref{fig4}(a)]. These values are significantly smaller than the pairing potential in the niobium electrode, $\Delta_{\rm Nb} \approx 1.3$~meV, which is estimated from a measurement of the critical temperature. It is worth pointing out that a similar suppression is found for the characteristic energy of the Josephson coupling in Josephson junctions with TI weak links. Typical values for the product of critical current and normal state resistance ($I_C R_N$) are 0.01 -- 0.2\,$\Delta_{\rm Nb}/e$, see table I in Ref.~\cite{Galletti2014}. The reduction is attributed to an interface barrier between the superconducting electrode and the TI material. Disorder effects must also be considered~\cite{Sau2010,*Potter2011}.

%These values set the relevant scale for the obtainable  product of the critical current and normal state resistance, $I_cR_N\leq\Delta'_{\rm STI}/e$, in topological Josephson junctions with these materials. Whether the reduced values of $\Delta'_{\rm STI}$ compared to the superconducting gap of Nb result from disorder or reflectivity of the Nb-TI interface ~\cite{Sau2010,*Potter2011} is not obvious from our experiment.} 

%\textcolor{dickred}{At even higher bias, $|V|>0.8$\,mV, the data further deviate below the $q=e$ fit both in zero and finite $B$-field [see Fig.~\ref{fig4}(a)]. We attribute this effect to a suppression of the shot noise by electron-phonon (\textit{e-ph}) energy relaxation~\cite{Nagaev1992, Blanter2000}. For $T_N\gtrsim5\,$K the \textit{e-ph} cooling dominates the noise response. This is consistent with two-dimensional (e.g., surface) acoustic phonons, see Supplementary Material for more details~\cite{supplemental}.}

%\cite{[{At even higher bias, $|V|>0.8$\,mV, the noise is influenced by electron-phonon relaxation, see Supplementary Material for more details~\cite{supplemental}}]remark_eph}

The observation of charge doubling and a $2e\rightarrow e$ transition in Figs.~\ref{fig3}(b) and~\ref{fig4}(a) is clear evidence of AR in $B=0$. Surprisingly, the Fano factor in the AR regime, $F_{\rm AR} = 0.22 \pm 0.02$, is considerably smaller than the universal  value, $F_{\rm N} = 1/3$, attained in the normal transport regime in magnetic field [Fig.~\ref{fig3}(c)] and in the reference N-TI-N device [Fig.~\ref{fig2}]. Similar behavior is found in all three devices with device resistances ranging from 1\,${\rm k\Omega}$ to 4\,${\rm k\Omega}$. This is our main observation. It is in sharp contrast to measurements in normal metal--~\cite{Kozhevnikov2000} and semiconductor--based~\cite{Choi2005} N-S contacts. In the former case, $F_{\rm AR}=F_{\rm N}=1/3$, as expected for the universal eigenvalue distribution of metallic diffusive conductors~\cite{Beenakker1997}, while in the latter case $F_{\rm AR}>F_{\rm N}$ was found~\cite{remark_Fano}.

%\textcolor{dickred}{Low energy in-gap density of states is known to reduce the shot noise, as is the case for $d$-wave and $p$-wave superconductor-based tunnel junctions~\cite{Zhu1999,Tanaka2000}. This effect can be understood intuitively in the framework of heat transport as follows, see Ref.~\cite{Nagaev2001}. In the case of perfect AR, the heat transport across the N-S interface vanishes, which results in exact shot noise doubling for diffusive NS junctions. The in-gap states can restore a finite heat conduction of the S-lead and cause the shot noise to reduce. In the presence of Majorana zero modes~\cite{Bolech2007,Fu2008,Tanaka2009,Zazunov2016}, the heat is carried away along the boundary of the superconductor by means of a gapless chiral Majorana mode. Yet, this is hardly relevant for the present experiment, since time reversal symmetry is obeyed~\cite{Alicea2012} in $B=0$. Most probably, the in-gap states emerge as a disorder effect in the bulk of a strongly proximized TI (STI). Thanks to the lateral geometry of our N-TI-S devices, this could result in a heat transport via STI, which occurs in parallel to the charge transport via Nb, explaining the reduction of $F_{\rm AR}$. The relevance of disorder is supported by the observed reduced values of $\Delta'_{\rm STI}$ compared to the superconducting gap of Nb. However, theoretically, the proximity effect in 3D TI is expected to be robust against disorder~\cite{Sau2010,*Potter2011}.  A further theoretical insight is clearly needed to explain our results on a microscopic level.} 

The observed reduction of the shot noise in the AR regime can be understood intuitively in the framework of heat transport as follows, see Ref.~\cite{Nagaev2001}. In the case of perfect AR, the heat transport across the N-S interface vanishes, which results in exact shot noise doubling for the N-S contact. Finite heat conduction in the S lead can reduce the shot noise below this limit. In particular, disorder effects may result in a finite, residual density of states in the strongly proximized TI (STI) below the Nb contact. This opens a parallel transport channel for heat conduction by quasiparticles thus reducing AR. Additionally, we consider the non-trivial gap structure of the induced proximity effect in the STI. It has been shown in~\cite{Zhu1999,Tanaka2000} that in-gap Andreev bound states which are formed in junctions with $p$- and $d$-wave superconductors due to a phase shift in the order parameter between different crystal directions alter the junction transmission and reduce the shot noise. We would like to mention that similar scenario has been proposed for N-FI-S structures where a finite magnetic polarization is induced by the ferromagnetic insulator (FI) and chiral Majorana zero mode forms~\cite{Tanaka2009}.

%In the case of perfect AR, the heat transport across the N-S interface vanishes, which results in exact shot noise doubling for diffusive NS junctions. A finite heat conduction of the S-lead causes the shot noise to reduce. In the presence of Majorana zero modes~\cite{Bolech2007,Fu2008,Tanaka2009,Zazunov2016} the heat can be carried away along the boundary of the superconductor by means of a gapless chiral Majorana mode. While this is hardly relevant for the present experiment, since time reversal symmetry is obeyed~\cite{Alicea2012} in $B=0$, we expect a similar effect owing to non-Majorana Andreev bound states, as is the case for $d$-wave and $p$-wave superconductor-based tunnel junctions~\cite{Zhu1999,Tanaka2000}. In addition, the in-gap states can emerge as a disorder effect in the bulk of a strongly proximized TI (STI). Thanks to the lateral geometry of our N-TI-S devices, this could result in a heat transport via STI, which occurs in parallel to the charge transport via Nb, explaining the reduction of $F_{\rm AR}$. A further theoretical insight is clearly needed to explain our results on a microscopic level.} 

%It has been recently suggested that in-gap states can also emerge as a disorder effect~\cite{Potter2010,*Potter2011} in the bulk of a strongly proximized TI (STI). Thanks to the lateral geometry of our N-TI-S devices, the heat transport via the STI occurs in parallel to the charge transport via Nb, which could explain the reduction of $F_{\rm AR}$. A further theoretical insight is clearly needed to explain our results on a microscopic level.

In summary, we investigated the shot noise in N-TI-S contacts defined on thin flakes of 3D TI $\rm Bi_{1.5}Sb_{0.5}Te_{1.7}Se_{1.3}$. We identified both the carrier charge $q$ and the Fano factor, responsible for the randomness of the discrete charge transport. In normal state, the elastic diffusive transport regime is evidenced via charge $q=e$ and nearly universal Fano factor $F_N\approx1/3$ in N-TI-S contacts in magnetic field and in a reference N-TI-N device. In the AR regime in $B=0$, the effective charge doubles, $q=2e$, whereas the Fano factor is considerably reduced $F_{\rm AR}\approx0.22\pm0.02$. Our main observation of $F_{\rm AR}<F_{\rm N}$, is in contrast with the available shot noise measurements in normal metal and semiconductor based devices~\cite{Kozhevnikov2000,Choi2005}.  Possible presence of low-energy Andreev bound states or in-gap states in the superconductor proximized 3D TI qualitatively explains our results.

We thank K.E.~Nagaev and Y.~Tanaka for fruitful discussions. This work was supported in part by the Russian Academy of Sciences, the Ministry of Education and Science of the Russian Federation Grant No. 14Y.26.31.0007, the RFBR Grants 16-32-00869 and  15-02-04285 and the Russian Science Foundation project no. 15-12-30030. Part of this work was funded by the Foundation for Fundamental Research on Matter (FOM), which is part of the Netherlands Organisation for Scientific Research (NWO), and the European Research Council.

%\bibliographystyle{unsrt}
%\bibliographystyle{apsrev}
%\bibliography{BSTSbbl}
%{\footnotesize\bibliography{InAsbbl}}

%

\end{document}

\title{Andreev reflection in s-type superconductor proximized 3D topological insulator. Supplemental Material.}
\author{E.S.~Tikhonov}
\affiliation{Institute of Solid State Physics, Russian Academy of
Sciences, 142432 Chernogolovka, Russian Federation}
\affiliation{Moscow Institute of Physics and Technology, Dolgoprudny, 141700 Russian Federation}
\author{D.V.~Shovkun} 
\affiliation{Institute of Solid State Physics, Russian Academy of
Sciences, 142432 Chernogolovka, Russian Federation}
\affiliation{Moscow Institute of Physics and Technology, Dolgoprudny, 141700 Russian Federation}
\author{V.S.~Khrapai} 
\affiliation{Institute of Solid State Physics, Russian Academy of
Sciences, 142432 Chernogolovka, Russian Federation}
\affiliation{Moscow Institute of Physics and Technology, Dolgoprudny, 141700 Russian Federation}
\author{M. Snelder}
\affiliation{MESA+ Institute for Nanotechnology, University of Twente, Enschede, the Netherlands}
\author{M.P. Stehno}
\affiliation{MESA+ Institute for Nanotechnology, University of Twente, Enschede, the Netherlands}
\author{Y. Huang}
\affiliation{Van der Waals - Zeeman institute, University of Amsterdam, the Netherlands.}
\author{M.S. Golden}
\affiliation{Van der Waals - Zeeman institute, University of Amsterdam, the Netherlands.}
\author{A.A. Golubov}
\affiliation{MESA+ Institute for Nanotechnology, University of Twente, Enschede, the Netherlands}
\affiliation{Moscow Institute of Physics and Technology, Dolgoprudny, 141700 Russian Federation}
\author{ A. Brinkman}
\affiliation{MESA+ Institute for Nanotechnology, University of Twente, Enschede, the Netherlands}

\maketitle

\section{Differential resistance in a wide bias range}
\begin{figure}[h]
\begin{center}
\vspace{10mm}
  \includegraphics[width=1\columnwidth]{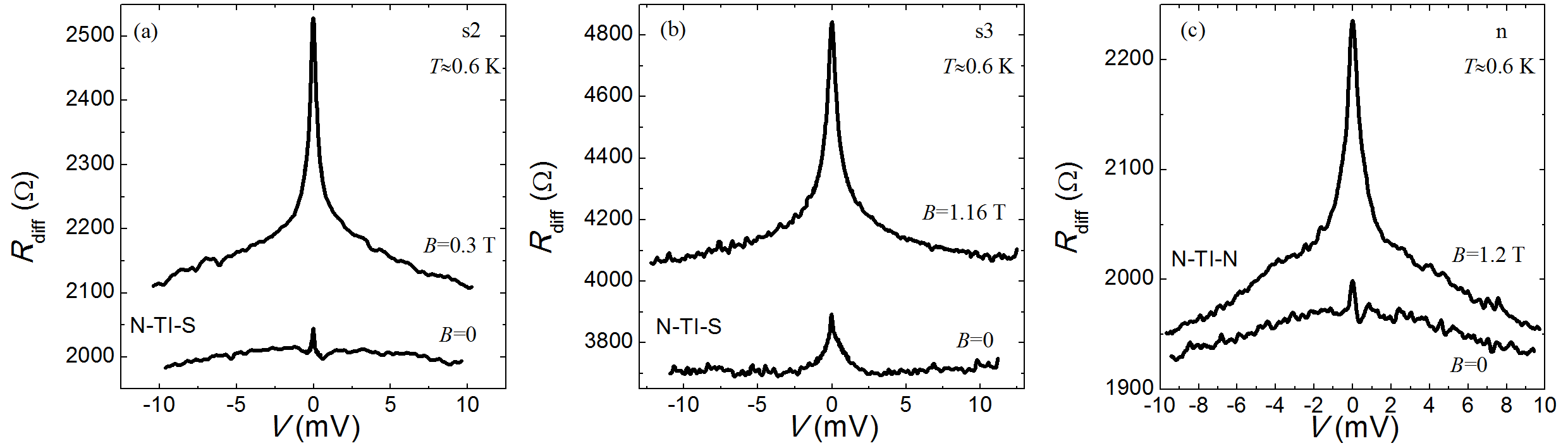}
   \end{center}
  \caption{Differential resistance in N-TI-S devices s2 (a) and s3 (b) and reference N-TI-N device n (c). The data is taken simultaneously with the main text noise data in Fig.~3 (s2), Fig.~4 (s3) and Fig.~2 (n).  
}\label{fig1}
\end{figure}

In all N-TI-S devices studied the differential resistance, $R_{\textrm{diff}}$, behaves similarly to the reference N-TI-N device and exhibits no AR related features. This is verified in Figs.~\ref{fig1}a and~\ref{fig1}b for two representative devices in a wide bias range. Just like in the reference N-TI-N device, see Fig.~\ref{fig1}c, the small zero bias feature in $B=0$ develops into a pronounced resistance peak in a magnetic field $B\sim1\,$T. This behavior is qualitatively consistent with a scenario of competing quantum corrections, weak anti-localization and Altshuler-Aronov, among which the former is suppressed by a perpendicular magnetic field and both are suppressed by a high bias owing to dephasing, see, e.g. Ref.~\cite{Lu2014}.

%\section{$2e\rightarrow e$ transition and proximity gaps}
%
%
%At increasing bias voltage in N-TI-S devices the noise temperature, $T_N$, demonstrates a change of the slope associated with a finite bias $2e\rightarrow e$ transition. As discussed in the main text, above the transition point the $T_N$ traces in zero and finite magnetic fields become identical up to a vertical shift. This is shown in Fig.~\ref{fig2} for all N-TI-S devices studied. The transition points, marked by arrows, measure proximity gaps in the range $0.23\,\textrm{meV}\leq\Delta\leq0.35$\,meV in different devices.
%
%
%\begin{figure}[h]
%\begin{center}
%\vspace{10mm}
  %\includegraphics[width=1\columnwidth]{fig2_supplemental.png}
   %\end{center}
  %\caption{Finite bias $2e\rightarrow e$ transition in N-TI-S devices. The traces in magnetic field are shifted by 0.14\,K, 0.19\,K and 0.32\,K, respectively for s1 (a), s2 (b) and s3 (c). The data is taken simultaneously with those in Fig.~3 of the main text (s1 and s2) and Fig.~4 of the main text (s3).		
%}\label{fig2}
%\end{figure}
\section{Electron-phonon energy relaxation}

\begin{figure}[h]
\begin{center}
\vspace{10mm}
  \includegraphics[width=1\columnwidth]{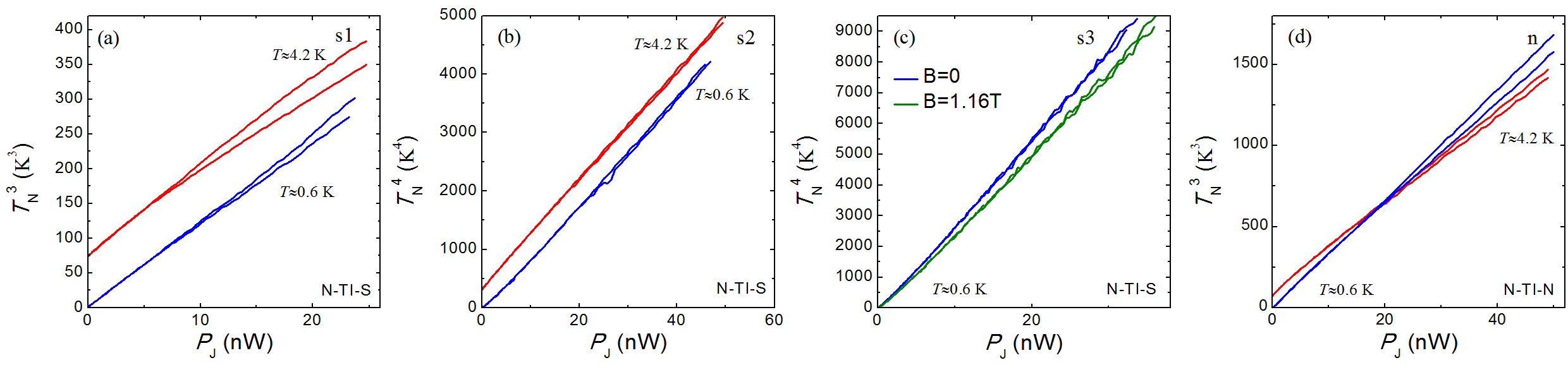}
   \end{center}
  \caption{E-ph energy relaxation in the strongly non-equilibrium transport regime. Close to linear dependence $T^{\alpha}_N\propto P_J$ at bath temperatures of $T=0.6$\,K (blue curves) and $T=4.2$\,K (red curves) in devices s1(a), s2(b), n(d) and at bath temperature of $T=0.6$\,K at zero (blue curve) and nonzero (green curve) magnetic field in device s3(c).		
}\label{fig2}
\end{figure}

As discussed in the main text, at large biases, $|V|>0.8$\,mV, the data deviate below the $q=e$ fit, both in zero and finite $B$-field, which is a result of shot noise suppression via electron-phonon (\textit{e-ph}) energy relaxation~\cite{Nagaev1992,Blanter2000}. We have checked that for $T_N>5\,$K the \textit{e-ph} cooling dominates the noise response and is consistent with the linear dependence $P_J\propto T_N^{\alpha}-T^{\alpha}$, where $P_J$ is the total dissipated Joule heat power and the exponent varies between ${\alpha}\approx3$ and ${\alpha}\approx4$ in different devices, see Fig.~\ref{fig2}. A cooling rate of this type might arise from the interaction with two-dimensional (e.g., surface) acoustic phonons~\cite{Kubakaddi2009,Viljas2010}, similar to graphene~\cite{Betz2012,Baker2012,Fong2012}, or the interplay of \textit{e-ph} and impurity scattering~\cite{Sergeev2000}. Note, that the doping dependence of the surface electrons' cooling rate in 3D TI~\cite{Herrero2012} $\rm Bi_{2}Se_{3}$ at much higher \textit{T} is consistent with the relaxation via surface acoustic phonons. 

\bibliography{supplemental}